\documentclass{article}
\usepackage{graphicx}
\PassOptionsToPackage{numbers, compress}{natbib}
\usepackage[preprint]{neurips_2026}

\usepackage[utf8]{inputenc} 
\usepackage[T1]{fontenc}    
\usepackage{hyperref}       
\usepackage{url}            
\usepackage{booktabs}       
\usepackage{amsfonts}       
\usepackage{nicefrac}       
\usepackage{microtype}      
\usepackage{xcolor}         
\usepackage{subfiles}
\usepackage{float}

\newcommand{\systemname}{\textsc{LEDGER}}
\newcommand{\systemfullname}{\systemname{} --- Layered Evidence and Decision Graphs for Execution Review}

\title{LEDGER: Claim-to-Evidence Trace Graphs for Auditing LLM Agents}

\author{%
    Daehong Kim \\
    Carnegie Mellon University \\
    Pittsburgh, PA 15213 \\
    \texttt{daehongk@andrew.cmu.edu} \\
    \And
    Haichao Miao \\
    Lawrence Livermore National Laboratory \\
    Livermore, CA 94550 \\
    \texttt{miao1@llnl.gov} \\
    \And
    Shusen Liu \\
    Lawrence Livermore National Laboratory \\
    Livermore, CA 94550 \\
    \texttt{liu42@llnl.gov} \\
}

\begin{document}

\maketitle

\begin{abstract}
    Large language model (LLM) agents can now carry out long-horizon technical workflows involving complex tool use, code execution, file edits, and generated artifacts. As agents do more work faster, the productivity bottleneck shifts from producing outputs to auditing whether those outputs are correct and trustworthy. Agent observability systems make fine-grained execution events visible, but visibility alone still leaves reviewers to reconstruct which actions, artifacts, and validation steps matter for a particular conclusion. We introduce \systemfullname{}, a tracing and review system that builds layered trace graphs over observed agent sessions. \systemname{} preserves Trace Records while grouping them into Evidence Nodes and Workflow Nodes, representing artifacts as evidence anchors, and adding typed semantic edges that connect claims to supporting actions, artifacts, and checks. Through data-analysis and coding examples, we show how the resulting traces expose workflow decisions, artifact lineage, repair steps, validation coverage, and claim-support paths for evidence-centered audit.

\end{abstract}

\section{Introduction}

Large language model (LLM) agents are increasingly used as interactive workers in technical and scientific workflows rather than only as single-turn question-answering systems \citep{ferrag2026survey, shin2025scientificworkflows}. A single session can include planning, data and code inspection, tool calls, code execution, file edits, and generated artifacts such as scripts, plots, tables, patches, and reports. As these sessions grow longer, their final results depend on a larger set of intermediate decisions, actions, and artifacts.

This changes where user effort is spent. An agent may quickly assemble a plausible analysis, patch, report, or conclusion, but accepting that result requires checking the work behind it. The user needs to know which inputs were inspected, which hypotheses or implementation choices were explored, which files were changed, which artifacts were produced, which checks were run, and which pieces of evidence support the final claim. In this setting, productive use depends not only on generation speed, but also on the user's ability to audit the recorded work.

Existing work on agent evaluation has recognized that an agent run should be assessed not only by its final output, but also by the sequence of actions that produced it \citep{yehudai2026evaluation, chen2025reasoningmodels, guan2026monitorability, wang2026trajaudit, li2026codetracer}. In particular, agent observability systems, e.g., langsmith \citep{langsmith2026}, make fine-grained execution records available to users, including prompts, tool calls, model responses, errors, intermediate outputs, and other detailed event-level records. 

Visibility, however, is not the same as auditability. An event trace can show what happened without showing which events matter for checking a particular conclusion. Human review is often evidence-centered: a reviewer starts from a reported claim and asks what supports it, which artifact contains the relevant evidence, what action produced or changed that artifact, and what check, if any, validated it. Answering these questions from a flat record requires reconstructing task structure and evidence relations by hand. The more carefully a user performs this reconstruction, the better the audit quality can be, but the cost grows with session length and complexity.

In this work, we introduce \systemfullname{}, which addresses this evidence-centered review problem by constructing layered trace graphs over observed agent sessions. \systemname{} does not replace observability; it builds on observed records and reorganizes them around audit needs. Instead of asking users to read an entire sequence of events, \systemname{} connects conclusions to the actions, artifacts, and checks that contributed to them, allowing a reviewer to move backward from a final claim to the concrete evidence behind it.

\systemname{} runs as a sidecar tracer alongside an interactive agent session. It captures session records, parses them into Trace Records, groups those records into lower-level Evidence Nodes, then groups related evidences into higher-level Workflow Nodes, and represents files, patches, command outputs, tables, plots, and other objects as artifact nodes.
Typed semantic edges such as \textit{uses}, \textit{produces}, \textit{checked\_by}, and \textit{supports} describe how actions, artifacts, validation steps, and claims relate to one another. 
The resulting graph makes the audit path explicit: a reviewer can start from a conclusion, follow support edges to relevant artifacts and actions, and inspect the source records behind them.
In addition, the graph is an aid to review, not a replacement for presenting source evidence. The proposed system provides direct indexing to the underlying artifacts, e.g., table / plots, for direct inspection. This lets reviewers inspect workflow-level structure while preserving access to the original records when details matter. The key contributions of the work are summarized as follows.
\begin{itemize}
    \item \textbf{A claim-to-evidence trace graph construction method.}
    We introduce a sidecar tracer that runs alongside an unmodified interactive agent session, parses captured messages, tool calls, outputs, file interactions, and artifacts into Trace Records, and organizes them into a layered semantic trace graph. Artifact nodes represent inspectable evidence objects, while typed edges connect claims to supporting work and validation steps.

    \item \textbf{An interface for evidence-centered agent review.}
    The interface exposes Workflow Nodes, Evidence Nodes, Trace Records, graph updates, artifacts, and tracer-agent call records in one dashboard. This lets reviewers move between graph-level audit paths and the source records behind them, including the records used to construct the graph itself.
    
\end{itemize}

\section{Related Work}

\subsection{Observability and Provenance for Agent Work}

Agent observability systems expose execution records such as prompts, tool calls, model responses, intermediate outputs, and errors \citep{langsmith2026}. Provenance work similarly records how data products and outputs are produced, from scientific workflow provenance and W3C PROV to recent ML and agent-workflow provenance systems \citep{davidson2008provenance, missier2013prov,padovani2025mlprovenance, souza2025provagent}. These systems provide the capture layer, whereas our focus is the review layer above it, i.e., arranging captured records around claims, artifacts, actions, and checks so a reviewer can traverse from a reported result to supporting evidence.

\subsection{Evaluating and Monitoring Agent Execution}

Agent evaluation increasingly treats the run itself as an object of assessment \citep{ferrag2026survey,yehudai2026evaluation}. ScienceAgentBench checks generated programs and execution results on scientific tasks \citep{chen2025scienceagentbench}; Graphectory computes process metrics over agent trajectories \citep{liu2026processcentric}; AgentAuditor scores safety and security behavior across steps \citep{luo2025agentauditor}; and monitorability work shows that monitor judgments depend on which parts of the run are visible \citep{guan2026monitorability}. These approaches motivate using intermediate actions as evidence, but they typically produce scores, metrics, monitor judgments, or failure labels. Our goal is instead focusing on supporting human review of agent output claim and the work / evidence behind it.
Explanation-faithfulness work reinforces the need to inspect execution evidence rather than relying only on generated explanations or final responses \citep{turpin2023language,chen2025reasoningmodels}. Because agent work interleaves reasoning and action \citep{yao2023react}, the audit record must reach the observed actions, artifacts, and checks. This also differs from agent orchestration graphs, whose edges define control flow before or during execution \citep{wang2024langgraph}; the proposed system builds a graph after capture to represent evidence relations for review.

\subsection{Interfaces for Evidence-Centered Trace Review}

Several systems make model reasoning or generated work inspectable through visual structure, including ReasonGraph, ReasonDiag, InteractiveKG, and DeepVIS \citep{li2025reasongraph,chen2026reasondiag,kim2026interactivekg,shuai2026deepvis}. Explanatory debugging frames the broader need for users to inspect system behavior when explanations and user understanding diverge \citep{kulesza2015explanatory}. These systems focus mainly on reasoning or generation traces, while execution review must also reach changed files, patches, command outputs, tables, plots, and other saved artifacts.
Graph of Trace is closest to our setting: it organizes scientific-agent execution events into a directed graph that updates while the agent works \citep{gao2026graphtrace}. \systemname{} differs in a few ways. It reads records emitted by an unmodified interactive session rather than relying on the working agent to self-report. It also allows interactive and direct traversal to the underlying artifact the model create, e.g., file, patch, table, plot, etc. 

\section{Building the Layered Trace Graph}

\systemname{} takes an observed agent session and constructs a layered trace graph for review. The graph is built from captured records and a session transcript, then organized into Trace Records, Evidence Nodes, Workflow Nodes, Artifact, and typed Semantic Edges. Capture and trace-record extraction preserve source records and provenance information. Graph construction then groups records into inspectable evidence units and workflow phases, extracts artifact anchors, and adds evidence relations as edges among actions, artifacts, checks, and claims.

\subsection{Capturing and Structuring Trace Records}
The lowest level of granularity for agent actions is a set of \textbf{Trace Records}: preserved observations of what happened during the session. Trace Records are not an interpretation of the agent's reasoning or task structure. They are the stable substrate from which the review graph is built, retaining message content, tool calls and results, lifecycle boundaries, artifact references, timing information, and links back to the original transcript. This separation matters because reviewers need to distinguish evidence that was captured from structure that was inferred during graph construction.
Our prototype tests this capture path on Codex using lifecycle hooks and transcript reconstruction, but the approach is not specific to Codex. Any coding agent can support the same style of trace graph if it exposes similar hooks, session transcripts and records for messages, tool invocations, tool results, file, and artifact references, etc

In the Codex implementation, the hook configuration records \texttt{SessionStart}, \texttt{UserPromptSubmit}, \texttt{PreToolUse}, \texttt{PostToolUse}, \texttt{PermissionRequest}, and \texttt{Stop} events. Each hook receives a JSON payload, appends that payload to a hook-event log, and, when a \texttt{transcript\_path} is available, refreshes a per-session dump of the current transcript. During a turn, these dumps are marked as live snapshots; after the \texttt{Stop} hook, the same files are refreshed as a final snapshot. The transcript dump contains session metadata, turn contexts, user and assistant messages, tool-call records, tool-result records, reasoning summaries when present, and lifecycle events. Hook records provide near-live boundary points, while the copied transcript is retained as the source of truth for complete reconstruction because hooks may not intercept every tool pathway or every non-shell event.
%
%

The hook path preserves two levels of evidence. First, it copies the Codex transcript JSONL unchanged. Second, it constructs a trace summary, \texttt{trace.json}, that normalizes transcript rows into a small set of record families: session and turn context, visible messages, tool calls and results, lifecycle events, reasoning summaries when present, parse errors, and coverage records. This step does not reinterpret the session semantically; it indexes transcript rows into stable record types while retaining links to the original transcript.
These Trace Records preserve the source context needed for review and rebuilding without forcing the reviewer to work directly from a transcript stream. Session and turn records capture the conditions under which the agent acted, message records preserve visible user and assistant communication, tool records bind invocations to their outputs, and coverage records make missing or unfamiliar transcript rows explicit rather than silently dropping them.
For graph construction, \systemname{} treats these Trace Records as the record layer beneath the semantic graph. Each run stores the hook payloads, transcript, and trace summary in a per-session directory. Downstream Evidence Nodes, Workflow Nodes, and edges can therefore retain source pointers into the evidence substrate, supporting later inspection and post-hoc rebuilding.

\subsection{Constructing Layered Review Structure}

The graph represents the logical flow of the session as relations among evidence, actions, artifacts, and claims. Review is rarely a straight reading of the log from top to bottom. A reviewer may start at a final claim and work backward to the evidence, or start from a changed file and ask which action produced it and whether it was checked.

\systemname{} builds the graph through three layers: \textbf{Trace Records}, \textbf{Evidence Nodes}, and \textbf{Workflow Nodes}. As discussed in the previous section, Trace Records preserve captured evidence. Evidence Nodes group those records into concrete, inspectable work units. Workflow Nodes group related Evidence Nodes into workflow phases. The layers are meant to keep the graph readable while still allowing a reviewer to return to the underlying records.
These layers are built by different means, and the difference sets how much trust a reviewer should have for each part of the graph. Trace Records come from deterministic transcript parsing and Trace Record extraction, which preserve captured-record order and links back to source records. The Evidence Layer and Workflow Nodes are produced by the tracer agent reading those records: which Trace Records form an Evidence Node, what type and category that node takes, which Evidence Nodes belong to the same Workflow Node, and which edges connect them. The Trace Records therefore stand on their own, while the structure built over them is an interpretation.


\begin{figure}[h!]
  \centering
  \includegraphics[width=0.85\linewidth]{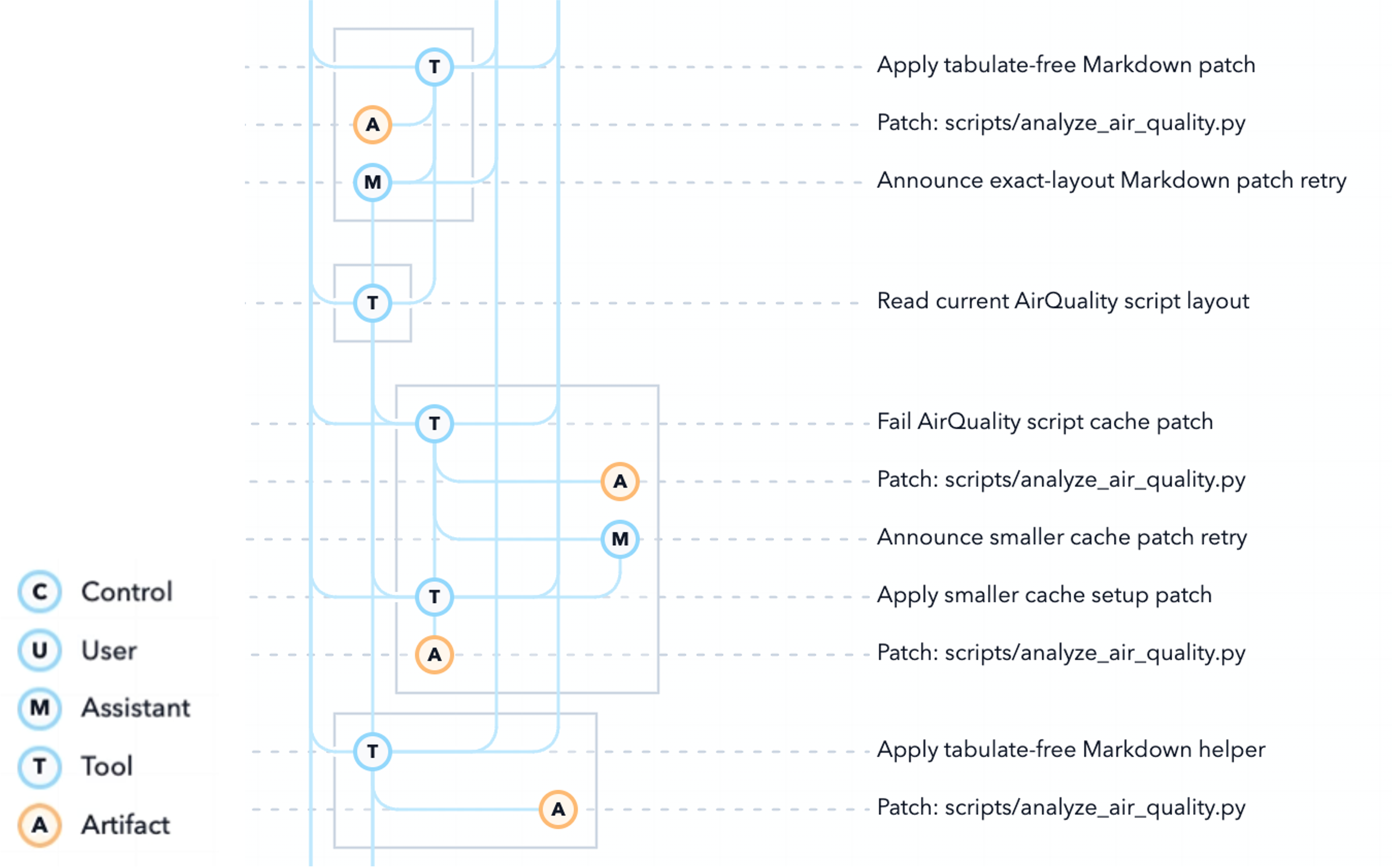}
  \caption{Evidence-node view from a housing recommendation data analysis session. The tabular dataset contains housing records, and the agent is tasked with recommending a house based on user preferences. Evidence Nodes indicate which files were touched, which methods were used to satisfy the goal, and how individual steps and results support one another. The legend for different type and category of evidence nodes are shown on the left.}
  \label{fig:evidence-node}
\end{figure}

An \textbf{Evidence Node} is the grouping above Trace Records. It turns closely related records into one inspectable unit of work. For example, a tool-call request and its result become one Evidence Node when they share a group id, which keeps a command and its output together. Figure~\ref{fig:evidence-node} illustrates this view for a housing recommendation task over a tabular air quality dataset: the nodes expose file touches, analysis methods, intermediate results, and the support relations among them. Evidence Nodes have a type and a category. Action nodes (blue) have different categories for indicating various session activity, while artifact nodes (orange) represent inspectable objects, e.g., a piece of code, a plot, as shown in Table~\ref{tab:low-categories}.

\begin{table}[th!]
  \caption{Evidence-node types and action categories. Action categories describe observable session activity; artifact nodes are separate Evidence Nodes for inspectable objects.}
  \label{tab:low-categories}
  \centering
  \begin{tabular}{lll}
    \toprule
    Type & Category & Definition / Description \\
    \midrule
    \textit{action} & \textit{control} & Session, turn, or task lifecycle record. \\
    \textit{action} & \textit{user\_message} & User request, correction, approval, or clarification. \\
    \textit{action} & \textit{assistant\_message} & Assistant plan, update, reasoning note, or answer. \\
    \textit{action} & \textit{tool\_call} & One tool invocation and its result, grouped by a shared id. \\
    \textit{artifact} & n/a & Inspectable object used, produced, or referenced by the agent. \\
    \bottomrule
  \end{tabular}
\end{table}


\begin{figure}[h!]
  \centering
  \includegraphics[width=0.75\linewidth]{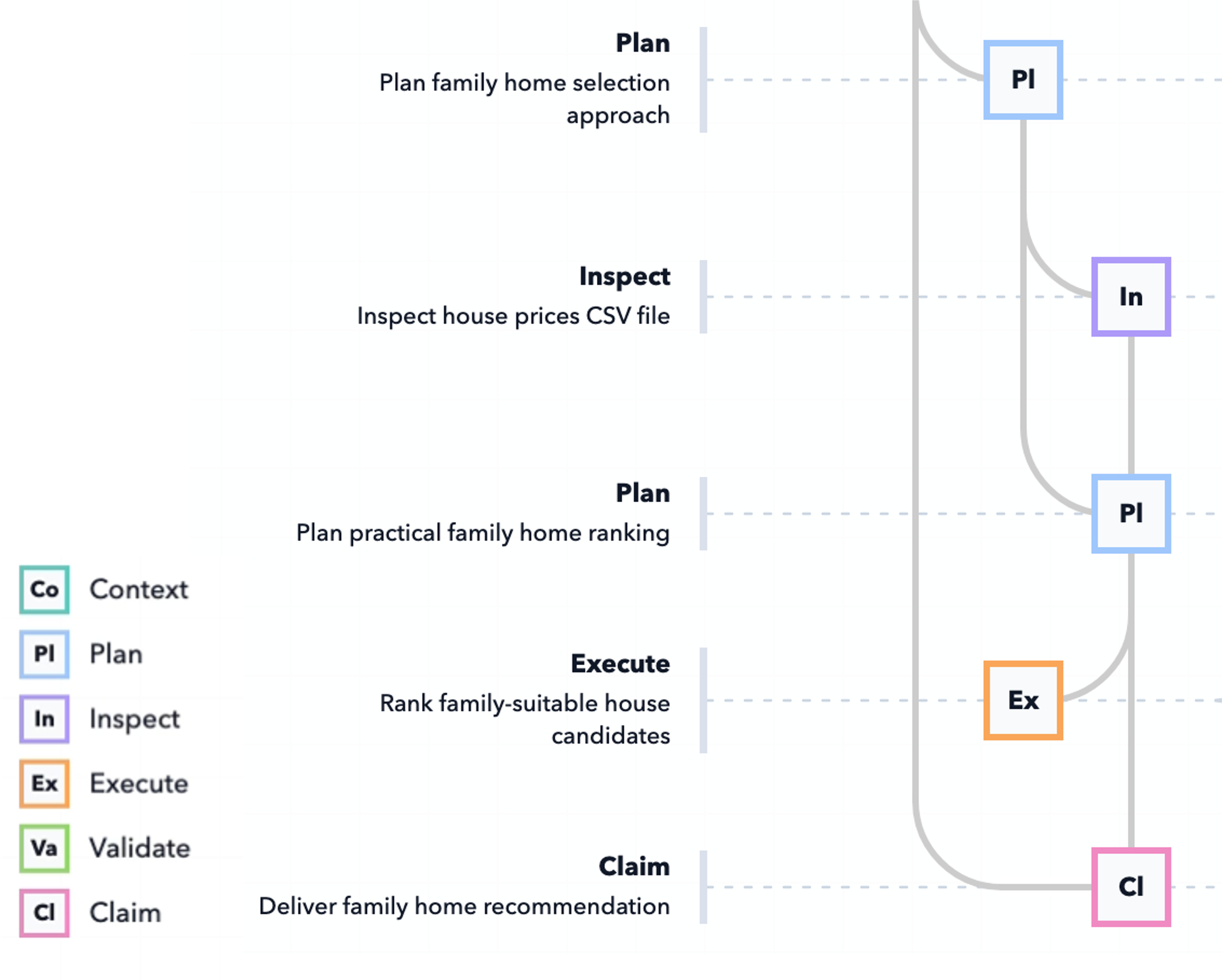}
  \caption{Workflow-node view. Workflow Nodes group Evidence Nodes into workflow phases so a long session can be reviewed at the phase level. The legend for different type of workflow nodes are shown on the left.}
\end{figure}

A \textbf{Workflow Node} explains how several Evidence Nodes work together in the task. This layer is needed because a session may contain many small actions that only make sense as part of a larger phase, such as inspecting data before analysis or validating a result before reporting it. The Workflow Node categories in Table~\ref{tab:high-categories} describe these phases.

\begin{table}[th!]
  \caption{Workflow-node categories. Workflow Nodes describe why Evidence Nodes belong together; the categories turn a long trace into task phases.}
  \label{tab:high-categories}
  \centering
  \begin{tabular}{ll}
    \toprule
    Category & Definition / Description \\
    \midrule
    \textit{context} & Frames the task, user goal, constraints, or session setting. \\
    \textit{plan} & Records the approach, scope, procedure, or tradeoff before work. \\
    \textit{inspect} & Gathers evidence by reading, searching, listing, previewing, or probing. \\
    \textit{execute} & Performs task-directed work such as editing, computation, analysis, or generation. \\
    \textit{validate} & Checks correctness or readiness through tests, builds, screenshots, or review. \\
    \textit{claim} & Records a final answer, explicit finding, status handoff, or user-facing conclusion. \\
    \bottomrule
  \end{tabular}
\end{table}

\subsection{Linking Artifacts and Audit Paths}

\textbf{Artifact nodes} are Evidence Nodes that represent objects inspectable after an action is over. A command output can be saved as evidence, a patch can change a file, and an analysis can create a plot or table. These objects are often the evidence that makes the work reviewable. Under hook tracing, artifact nodes are recovered from tool-call Trace Records and their results. The command, its arguments, and its output carry the file path and the kind of interaction, and the tracer reads them from that text.

When an object is represented as an artifact node in the Evidence Layer, the graph can connect Workflow Node claims to concrete evidence. A reviewer can see which Evidence Node produced an artifact, which later Evidence Node used it, and whether a Workflow Node containing a final claim is supported by that artifact. This matters most when the claim depends on an object outside the message stream, such as a generated figure, a changed source file, or a saved result table.

\textbf{Edges} encode how one graph element matters to another, and they operate at both graph levels. In the Evidence Layer, edges connect action Evidence Nodes and artifact Evidence Nodes, such as a command using a file, a command producing output, or a test checking a changed file. At the workflow level, edges connect Workflow Nodes when one phase frames, informs, checks, or supports another phase or claim. Cross-layer edges keep the summary accountable by linking each Workflow Node back to the Evidence Nodes and Trace Records that support it. Edges follow trace order, but they are added only when there is a semantic relation.
These edges support review. They let a reviewer move from a final claim to supporting work, or from an artifact to the action that produced it. The graph is therefore a map of evidence relations on top of a visual summary of the session.

\begin{table}[th!]
  \caption{Edge types. Edges are directed in trace order, and each type names the semantic relation that makes the source and target relevant to each other.}
  \label{tab:edge-types}
  \centering
  \begin{tabular}{ll}
    \toprule
    Type & Relation definition \\
    \midrule
    \textit{frames} & The source requirement, constraint, or plan sets context for the target. \\
    \textit{uses} & A work unit uses an artifact or resource as input. \\
    \textit{produces} & A work unit creates or changes an artifact, output, or state. \\
    \textit{informs} & The source evidence or result shapes the target analysis, plan, or action. \\
    \textit{checked\_by} & The source change or result is checked by the target validation step. \\
    \textit{supports} & The source evidence or finding justifies the target claim or committed choice. \\
    \bottomrule
  \end{tabular}
\end{table}

\section{LEDGER Interface}

\begin{figure}[h!]
  \centering
  \includegraphics[width=0.95 \linewidth]{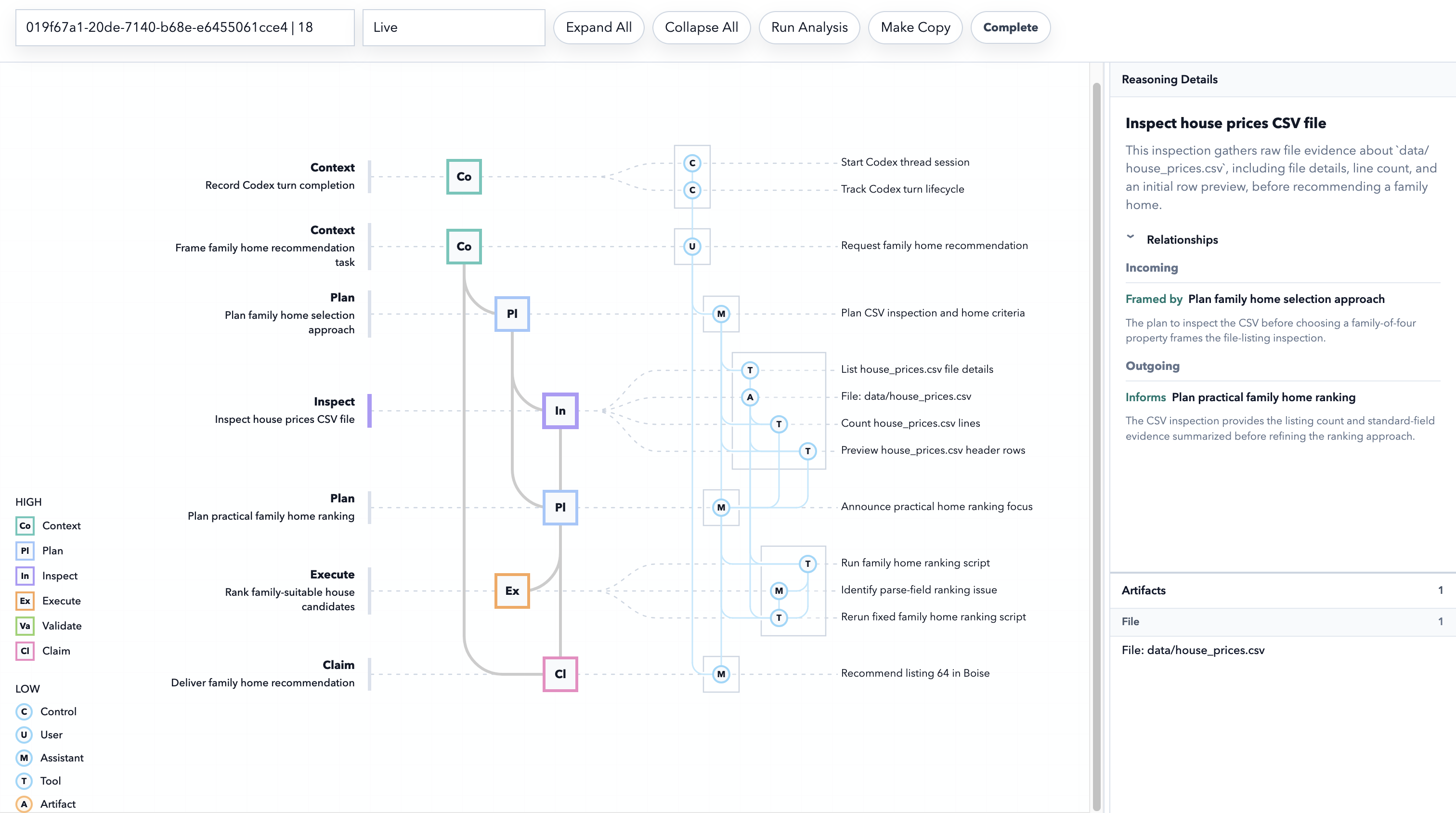}
  \caption{Review interface. The interface combines graph views, node details, artifacts, and event logs so a reviewer can move between workflow structure and source evidence.}
\end{figure}

\textbf{Graph-guided workflow review.} \systemname{} includes a local dashboard for reviewing traced sessions. The interface starts with a two-level graph which allows a compact view of session phases and evidence relations. Workflow Nodes summarize workflow phases, while Evidence Nodes show the concrete messages, tool calls, outputs, files, patches, and artifacts under those phases. A reviewer can first use the two-level graph to check whether the workflow covers the requested work, relevant inspection steps, validation steps, and final claim. The reviewer can then move into the Evidence Layer to inspect the specific work units that support or weaken that interpretation.

\begin{figure}
    \centering
    \includegraphics[width=0.95\linewidth]{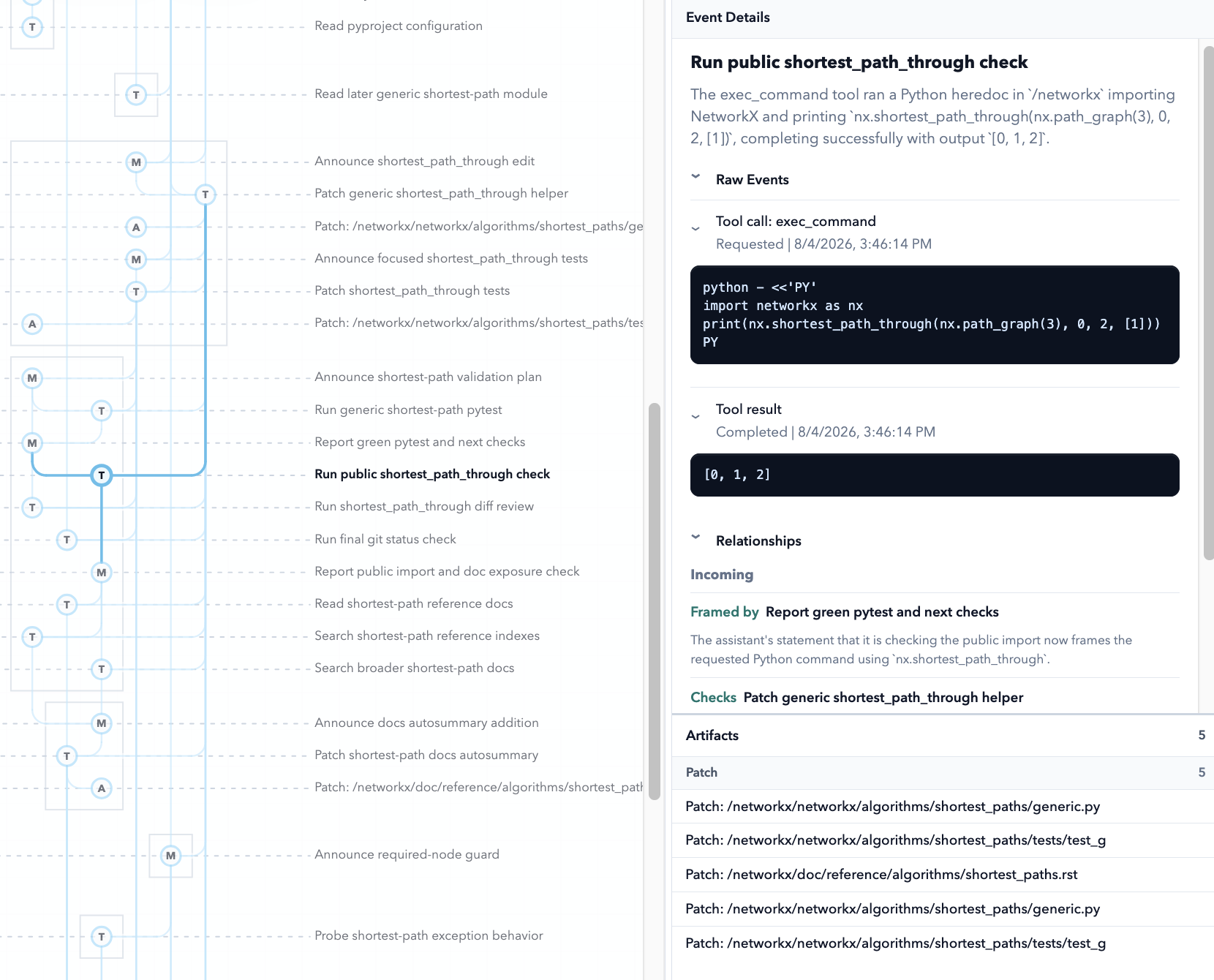}
    \caption{Event-detail view for inspecting evidence. Selecting an Evidence Node opens the recorded tool call, tool result, semantic relations, and related artifacts, allowing the reviewer to move from a event node to the source records behind it.}
    \label{fig:event-detail-view}
\end{figure}

\textbf{Evidence inspection.} Node details and artifact views connect the graph back to evidence. Selecting a node shows its category, status, description, evidence links, Trace Record links, and related artifacts. Artifact views let a reviewer inspect saved terminal output, patches, referenced files, generated plots, result tables, and other saved evidence. Review does not stop at the graph. The graph is an index into the recorded work. A final claim can be checked by following its supporting edges to the work that produced it, then opening the artifacts and Trace Records behind that work.

\textbf{Trace-construction audit.} Event and log views show Trace Records, graph updates, Evidence Layer edges, tracer events, model-call records, and trace errors. They make graph construction itself reviewable, so a reviewer can check how captured records became Trace Records and how Trace Records were grouped into Evidence Nodes and Workflow Nodes. When the graph appears incomplete, these views help a reviewer distinguish an agent error from a tracing error. By keeping the graph, artifacts, Trace Records, and trace logs together, the interface supports review of both the agent's work and the trace built from that work.

\section{Case Studies}

We use two examples to showcase how the trace graph and dashboard support review across different kind of agent workflows. The analysis focuses on how a live trace presents workflow phases, important decisions, artifacts, checks, and claims to a human reviewer.
The first example is a data-analysis use case that highlight artifact lineage, where a reported finding traces back through tables and a plot to the source data. The second coding example illustrate code grounding, where a change traces back to the repository reading that placed it and forward to the tests that checked it. Both sessions were run in Codex using GPT-5.5 with live tracing enabled, then reviewed through the dashboard.

\subsection{Case Study 1: Tabular Data Analysis}
The data-analysis case study uses an hourly air-quality dataset with sensor readings, reference pollutant measurements, timestamps, and weather variables. This kind of data-driven scientific workflow is a common setting for agent evaluation because the run includes code, data cleaning, execution results, and cost \citep{chen2025scienceagentbench}. The task asks the agent to clean the dataset and characterize the typical daily pattern of the main pollutant measurements. The full prompt is included in Appendix~\ref{app:case-prompts-1}.

The task gives the trace several checkable points because parsing and cleaning errors would change the hourly summary. The file requires timestamp parsing, numeric conversion, and handling of its documented missing-value convention, which is prone to error.

The expected trace should show a focused data-analysis flow: file inspection, parsing, missing-value handling, column selection, hourly aggregation, plot generation, and report writing. The produced artifacts can be checked directly. A reviewer can reopen the missingness table, hourly summary table, and plot, then compare them against the final report.


\begin{figure}
    \centering
    \includegraphics[width=0.99\linewidth]{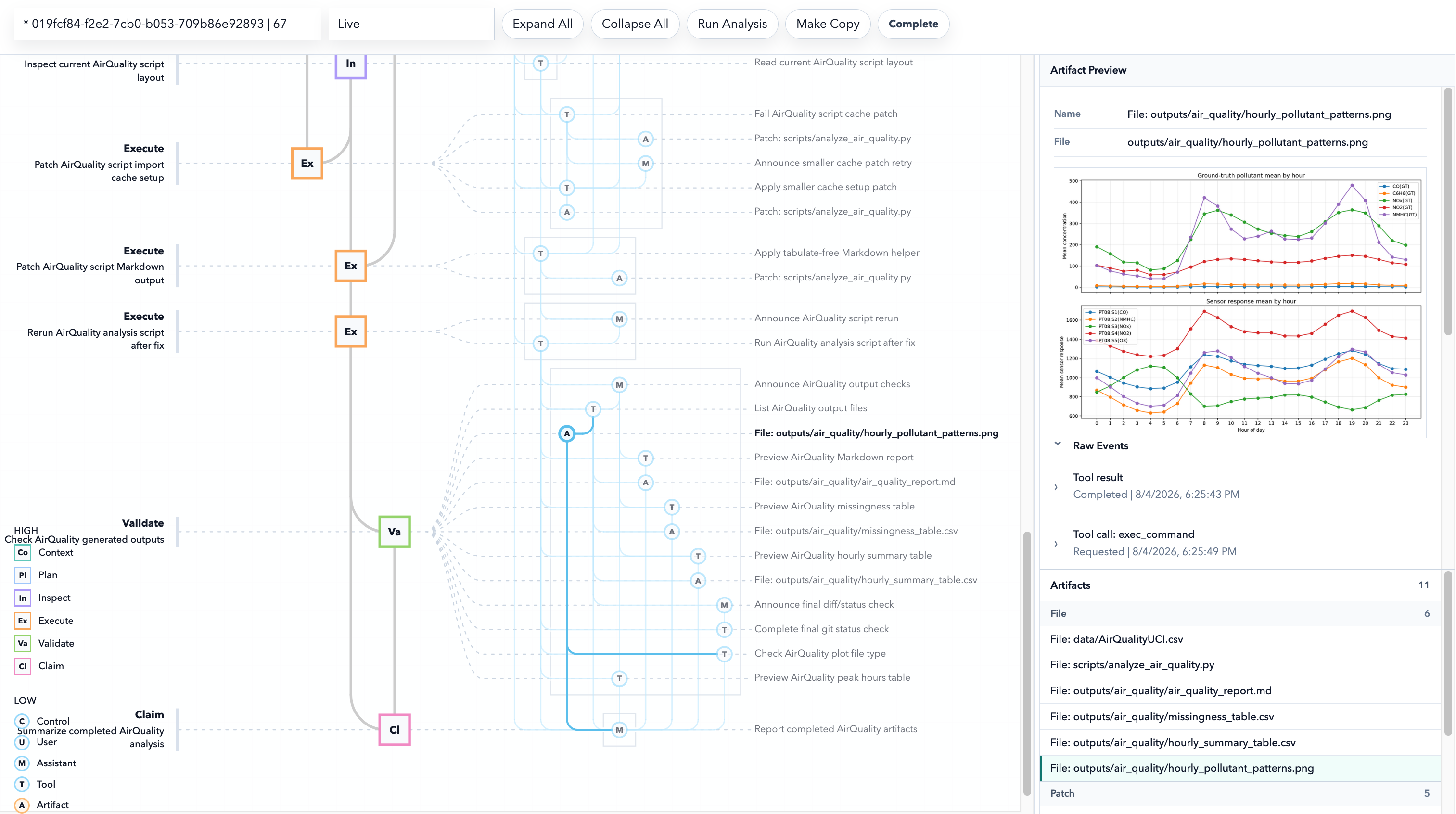}
    \caption{Artifact-centered review path in the data-analysis case study. The selected plot artifact is shown with the surrounding output-checking phase, linking the generated figure, tables, report, and final claim back to the reported analysis steps.}
    \label{fig:case1-artifact}
\end{figure}

\textbf{Results.} The data-analysis trace emphasized a path from source data to reported findings. As illustrated in Figure \ref{fig:case1-artifact}, in the Workflow Node view, the cleaning and summary plan appeared after file inspection and before script generation, so the reviewer could see that the analysis choices were grounded in the observed file format. Selecting those phases in the interface exposed the CSV preview, row-count check, ending-row inspection, and environment checks that preceded the script.

The trace also exposed an error-and-repair sequence. The first script execution failed because the generated script used a Markdown-table call that required an unavailable optional dependency. The Evidence Layer linked the failed command output and traced back to the later repair steps, including script patches, a successful rerun, and validation of the generated report, tables, and plot.

The central review path was the output-checking phase. The dashboard linked the final claim to the generated report, tables, and plot, and then back to the script creation, failed execution, repair patch, successful rerun, and preview commands that produced or checked those artifacts. This made the reported daily pattern inspectable as a chain from the final claim to the artifact and then to the recorded process behind it. The same path also placed quality-check artifacts, such as the missingness table, next to the reported results, so a reviewer could inspect whether the report was supported by the available data.

\begin{figure}
    \centering
    \includegraphics[width=0.99\linewidth]{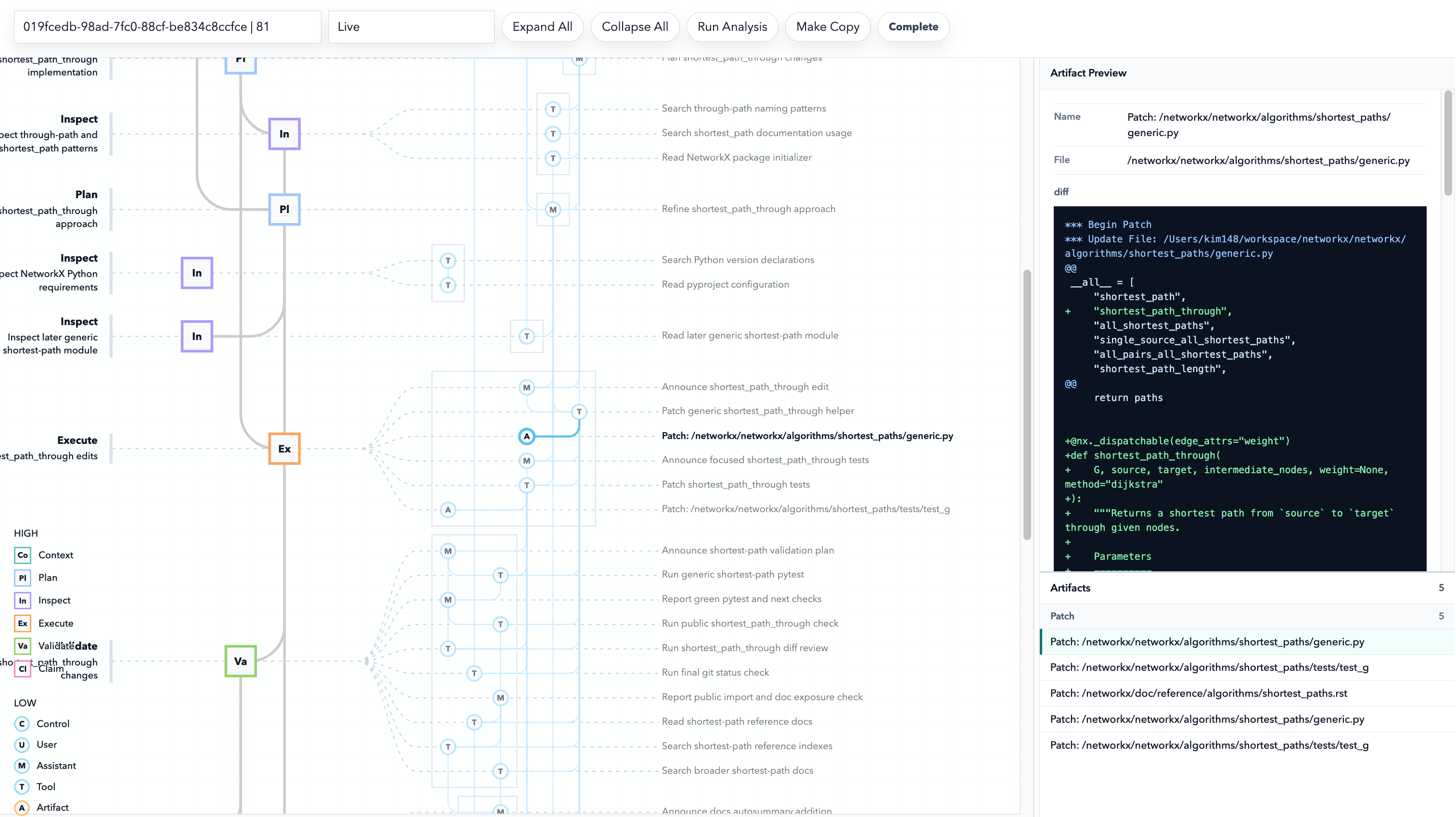}
    \caption{Patch-centered review path in the coding case study. The dashboard links the implementation patch to the surrounding inspection, implementation, and validation steps, allowing the reviewer to inspect the code change alongside its supporting records.}
    \label{fig:case2-patch}
\end{figure}

\subsection{Case Study 2: Feature Addition in an Existing Codebase}

The coding case study uses NetworkX, an established Python graph library, as an existing codebase \citep{hagberg2008networkx}. It asks the agent to add a shortest-path utility that returns a path from a source to a target while visiting a given sequence of intermediate nodes in order. The full prompt is included in Appendix~\ref{app:case-prompts-2}.

The agent must first understand the existing software structure. It needs to inspect where shortest-path functions live, how NetworkX exposes algorithm functions, how weighted and unweighted paths are handled, and how nearby code raises errors for missing nodes or missing paths. A correct answer therefore depends on reading the codebase as well as writing new code.

The expected trace should show a typical feature-development flow: repository inspection, function placement, implementation, test addition, documentation exposure, and validation. The trace should make it possible to check which files the agent read before editing, whether the new function follows nearby patterns, which tests were added, and whether the tests were run. If the implementation solves a harder unordered-waypoint problem, treats the result as a globally simple path, or handles missing paths in a way that does not fit the project, the relevant commands, edits, and test results should appear in the recorded work.


\textbf{Results.} The trace separated the agent-designed implementation from the part it repaired. The first implementation passed its test run, and only afterward did the trace add a check of how existing shortest-path functions handle missing nodes, followed by a guard patch, a regression-test patch, and a second focused test run. In the final diff these two stages are one function. In the trace they are separate phases with separate evidence, so a reviewer can tell which behavior came from the design and which came from an edge case the agent raised after the fact.
 The same separation fixed what the completion claim covers. The first test run checked the ordered-visit behavior on weighted and unweighted graphs, and the regression test checked the guard. The final status reports one passing run, while the trace records which check covered which behavior. A reviewer can therefore read the extent of the support behind the claim rather than only that support exists.
 
Before the edit, the Workflow Node view separated repository inspection into searches for shortest-path exports, reads of the generic shortest-path module, reads of nearby tests, and checks of documentation lists. These phases opened into the file reads and search results that led the agent to place the new utility in the generic shortest-path module, and they linked the implementation patch to the prompt requirement that the path visit intermediate nodes in order and to the shortest-path functions it reused. The design choice was therefore checkable rather than assumed: ordered segment composition over existing shortest-path calls.
The final claim traversed back to this evidence through its \textit{supports} edges, which reached the changed files, the documentation update, the focused test run, the import check, and the final diff review.

\section{Discussion}

Agent sessions are usually reviewed as a stream of messages, where a reader follows the session forward and reconstructs the connection between a reported result and the work behind it. This paper described \systemname{}, which organizes the same session as a graph of evidence relations instead. A reviewer can start from a claim, follow supporting edges to the actions and artifacts behind it, and open the Trace Records under those graph elements. The trace turns the link between a claim and its evidence into something to traversable rather than something need to be reconstructed by the user.

The two case studies exercised different review paths. In the data-analysis task, the trace carried artifact lineage from the source file through cleaning steps, generated tables, and plots to the reported daily pattern. In the coding task, it carried code grounding from repository inspection through patches and tests to a completion claim. Both paths rest on the same design principle: Trace Records preserve what happened, while Evidence Nodes, Workflow Nodes, artifact anchors, and semantic edges make those records usable as review objects.

The main limitation is that the current graph above the Trace Record layer is not fully deterministic. Captured hook payloads, copied transcripts, Trace Records, and coverage summaries provide deterministic anchors, but the tracer still interprets which records belong together, which Workflow Node they express, and which semantic edge best describes a relation. These choices can be incomplete, unstable, or wrong. \systemname{} therefore treats the graph as an audit aid rather than a source of truth: the interface keeps the underlying records visible so reviewers can check the graph construction itself. This limitation is closely related to concerns in reasoning faithfulness, monitorability, and provenance \citep{chen2025reasoningmodels,guan2026monitorability,souza2025provagent}.

Future work should make trace construction more robust by replacing model-inferred structure with deterministic or independently verifiable structure wherever possible. File reads, writes, patches, command outputs, test runs, and generated artifacts can be tracked through stronger instrumentation, structured tool records, or data-provenance backend \cite{devarajan2024dftracer}. For data-analysis workflows in particular, provenance-aware backends could expose table-, column-, and transformation-level dependencies rather than requiring the trace builder to recover them from command text and output previews. Repeated-run studies should also measure whether the same session produces stable nodes, edges, and claim-support paths under reconstruction.

There is also room to improve the graph vocabulary and visual design. Edge types such as \textit{uses}, \textit{informs}, \textit{supports}, and \textit{checked\_by} are useful for review, but their boundaries can blur in long sessions. Future systems should distinguish temporal, dataflow, validation, and evidential relations more explicitly, expose how each edge was produced, and encode uncertainty or missing support directly in the graph. Better visual encodings could separate deterministic from inferred structure, emphasize unsupported claims and unchecked artifacts, and reduce clutter by showing claim-centered or artifact-centered subgraphs by default.

The broader goal remains the same: as agents produce more work, review systems should help users audit the connection between conclusions and the work that produced them. Layered trace graphs are one step toward that goal because they preserve access to source records while making actions, artifacts, checks, and claims inspectable as connected evidence.

\begin{ack}
This work was performed under the auspices of the U.S.\ Department of Energy (DOE) by Lawrence Livermore National Laboratory under Contract No. DE-AC52-07NA27344. This work was primarily supported by DOE ECRP 51917/SCW1885. 
This work is reviewed and released under LLNL-CONF-2022801.
\end{ack}

{\small
\bibliographystyle{unsrtnat}
\bibliography{references}
}


\appendix


\section{Case Study Prompts}

\subsection{Case Study 1: Tabular Data Analysis Prompt}
\label{app:case-prompts-1}

\begin{quote}
\small
Analyze the hourly air-quality dataset in \texttt{data/airqualityuci.csv} as a reproducible tabular data analysis task.

First inspect and clean the dataset. Parse the date and time fields into timestamps, convert numeric columns correctly, and handle the documented missing-value convention. Report the number of rows, the date range, and missingness by column.

Then characterize the typical daily pattern for the main pollutant measurements. Compute the mean value by hour of day, identify the peak hour for each pollutant, and support the answer with one summary table and one plot.

Produce a short Markdown report, a missingness table, an hourly summary table, and one plot.
\end{quote}

\subsection{Case Study 2: Feature Addition Prompt}
\label{app:case-prompts-2}

\begin{quote}
\small

Look at the NetworkX repository.

Add a utility function that returns a shortest path from a source to a target while visiting a given sequence of intermediate nodes in order.

The function should build on existing shortest path functions, support weighted and unweighted graphs, concatenate segment paths without repeating boundary nodes, and follow existing NetworkX conventions for missing nodes or missing paths. Treat the result as a path or walk through the required nodes, not as a globally simple path constraint.

\end{quote}

\newpage

\section{Case Study Full Trace Graphs}

\subsection{Case Study 1 Trace Graph}
\begin{figure}[h!]
    \centering
    \includegraphics[height=0.8\textheight]{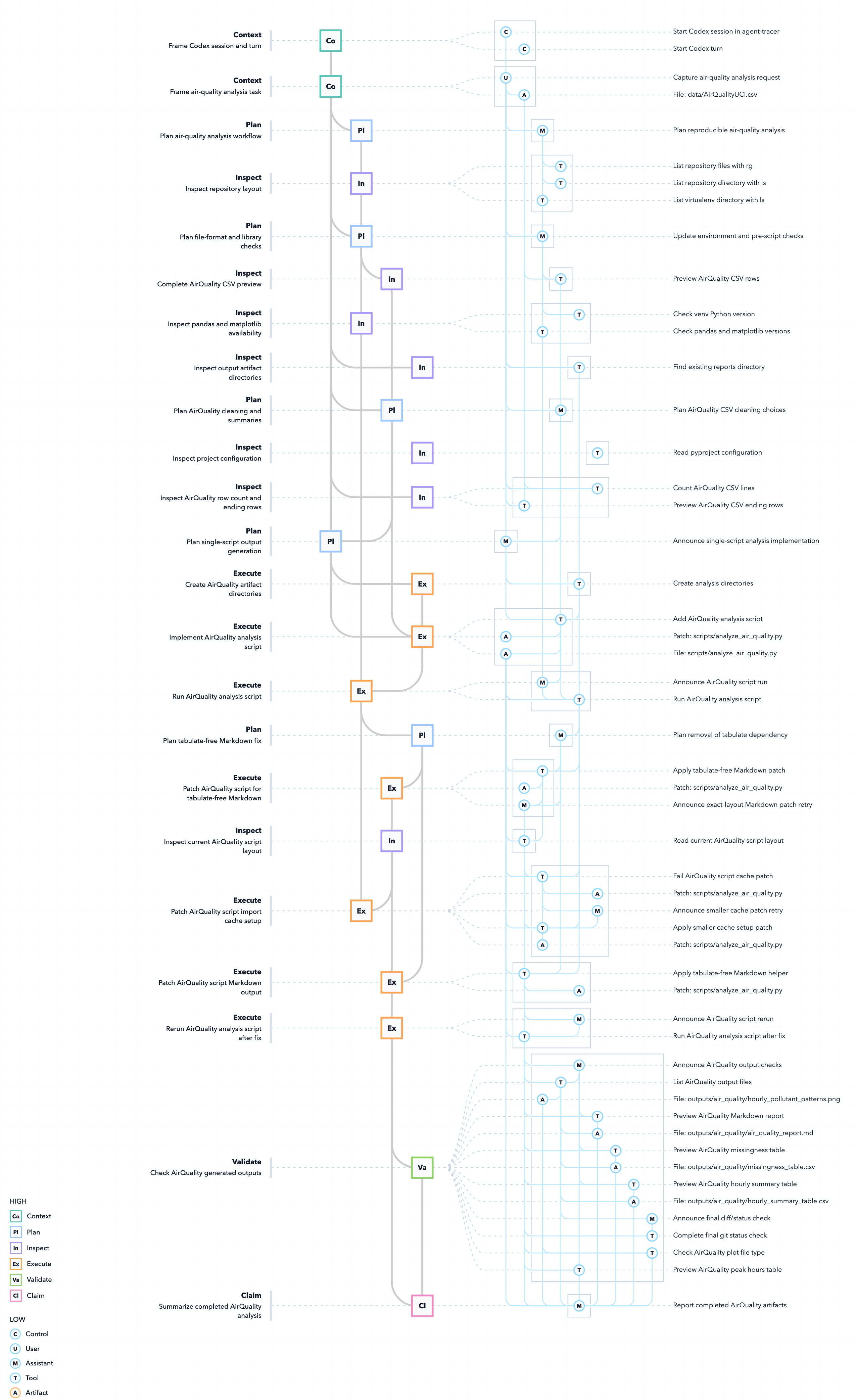}
    \caption{Case Study 1 Trace Graph}
    \label{fig:case1_graph}
\end{figure}

\newpage
\subsection{Case Study 2 Trace Graph}
\begin{figure}[h!]
    \centering
    \includegraphics[height=0.8\textheight]{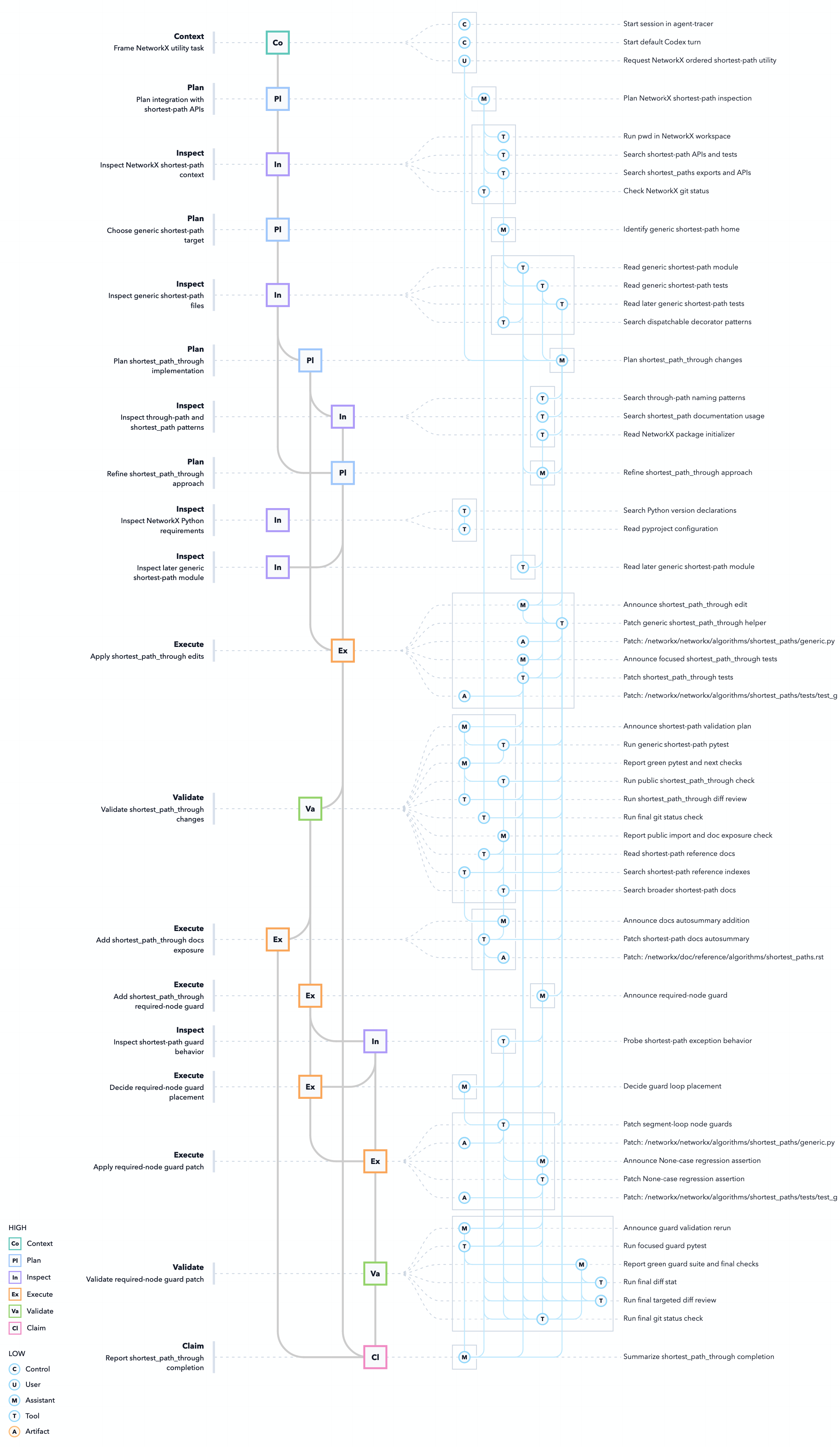}
    \caption{Case Study 2 Trace Graph}
    \label{fig:case2_graph}
\end{figure}


\newpage

\end{document}